\documentclass{emulateapj}
\usepackage{natbib}
\usepackage{xspace}
\usepackage{amssymb}
\usepackage{amsmath}
\usepackage{graphicx}

\def\revised#1{{#1}}
\def\aap{A\& A}

\def\mnras{MNRAS}
\def\nat{Nature}
\def\apj{ApJ}

\def\apjl{ApJL}
\def\apjs{ApJS}

\def\sec{\hbox{s}}

\def\comma{\,,}

\def\rcentr{R_{\mathrm{centr}}}

\begin{document}

\shorttitle{Crystalline silicates and disk accretion history}
\shortauthors{Dullemond, Apai \& Walch}
\title{Crystalline silicates as a probe of disk formation history}
\author{C.P.~Dullemond}
\affil{Max Planck Institut f\"ur Astronomie, K\"onigstuhl 17,
D69117 Heidelberg, Germany; Email: dullemon@mpia.de}
\author{D.~Apai}
\affil{Steward Observatory, 933 N. Cherry Avenue, Tucson AZ 85721, USA; 
NASA Astrobiology Institute}
\author{S.~Walch}
\affil{Uni-Sternwarte M\"unchen, Scheinerstr. 1, D--81679 M\"unchen, Germany}

\begin{abstract}
%
We present a new perspective on the crystallinity of dust in protoplanetary
disks. The dominant crystallization by thermal annealing happens in the very
early phases of disk formation and evolution. Both the disk properties and
the level of crystallinity are thereby directly linked to the properties of
the molecular cloud core from which the star+disk system was formed. We show that, under
the assumption of single star formation, rapidly rotating clouds produce
disks which, after the main infall phase (i.e.~in the optically revealed
class II phase), are rather massive and have a high accretion rate but low
crystallinity. Slowly rotating clouds, on the other hand, produce less
massive disks with lower accretion rate, but high levels of
crystallinity. Cloud fragmentation and the formation of multiple stars
complicates the problem and necessitates further study. The underlying
physics of the model is insufficiently understood to provide the precise
relationship between crystallinity, disk mass and accretion rate. But the
fact that with `standard' input physics the model produces disks which, in
comparison to observations, appear to have {\em either} too high levels of
crystallinity {\em or} too high disk masses, demonstrates that the
comparison of these models to observations can place strong contraints on
the disk physics. The question to ask is not why some sources are so
crystalline, but why some other sources have such a {\em low} level of
crystallinity.
%
\end{abstract}




\section{Introduction}
One of the major discoveries achieved with infrared spectroscopic studies of
Herbig Ae/Be stars and classical T Tauri stars is that a significant
percentage of the dust in their circumstellar disks is of crystalline form
(e.g.~Bouwman et al.~\citeyear{bouwmanmeeus:2001}; van Boekel et
al.~\citeyear{vboekelmin:2005}; Apai et al.~\citeyear{apaiscience:2005}).
Also in our own solar system there is evidence, from observations of comets,
that the dust of the primordial solar nebula was partly crystalline (Wooden
et al.~\citeyear{woodenharker:1999}). This poses an interesting puzzle
because the dust inherited from the interstellar medium must have been
amorphous (Kemper et al.~\citeyear{kempervriend:2004,kempervriend:2005}) and
hence the crystallization must have taken place {\em within} the disk. Since
silicate dust crystallizes only when heated to relatively high temperatures
($\gtrsim$ 800 K), the discovery of these crystals proved that
the dust in such disks is (or has been) subject to processes involving high
temperatures.

The very inner regions of protoplanetary disks (i.e.~inward of the
`crystallization radius' which is about 0.1 AU for a T Tauri star and 0.7 AU
for a Herbig star) are warm enough to thermally anneal the dust, and thus
produce crystalline silicates like enstatite and forsterite.  But the
presence of crystalline dust in Solar System comets proves that processed dust
also existed beyond the snow line where comets formed, well outside the
crystallization radius (e.g.~Wooden et al.~\citeyear{woodenharker:1999}; Min
et al.~\citeyear{minhov:2005}). Similarly, the infrared spectra of T Tauri
and Herbig \revised{disks} show that crystalline silicates often exist out to radii as
cool as 150 K. And by directly spatially resolving the disk down to 2 AU
with interferometric 10 $\mu$m spectroscopy it has been shown that for many
sources the outer cooler disk regions clearly contain crystalline dust (van
Boekel et al.~\citeyear{vanboekelnature:2004}), \revised{though a smaller 
fraction than in the inner disk regions.}

Currently the most favored theory involves radial mixing (Gail
\citeyear{gail:2001,gail:2002}; Wehrstedt \& Gail \citeyear{wehrgail:2002};
Bockel\'ee-Morvan et al.~\citeyear{bockmorvan:2002}, based on earlier ideas
by Morfill \& Voelk \citeyear{morfillvoelk:1984}). Even though the main
accretion stream in a disk points inward toward the star, the turbulent
mixing within the disk can transfer at least a small amount of the thermally
processed dust grains to larger radii. This typically results in decrease of
crystalline silicate abundance with radius beyond the point where the
temperature drops below the annealing temperature. However, this decline is
slow enough that still a measureable quantity of crystalline silicates exists
at radii where the 10 $\mu$m silicate feature is produced in protoplanetary
disks, and where comets were formed in the protosolar nebula
(Bockel\'ee-Morvan et al.~\citeyear{bockmorvan:2002}). Whether \revised{or
not} thermal
annealing and radial mixing alone are sufficient to explain the levels of
crystallinity typically observed is debated. A possibility to enhance the
outward transport of crystalline silicates was recently discussed by Keller
\& Gail (\citeyear{kellergail:2004}), who have shown that accretion in the
inner disk regions is not always inward: While the surface layers of a disk
tend to accrete inward, the midplane of the disk moves outward. Any
thermally processed grains may be transported very efficiently outward with
this mechanism. Another possibility is the outward transportation of dust
via wind (Shu et al.~\citeyear{shushang:1996}).

\begin{figure*}
\centerline{
\includegraphics[width=5cm]{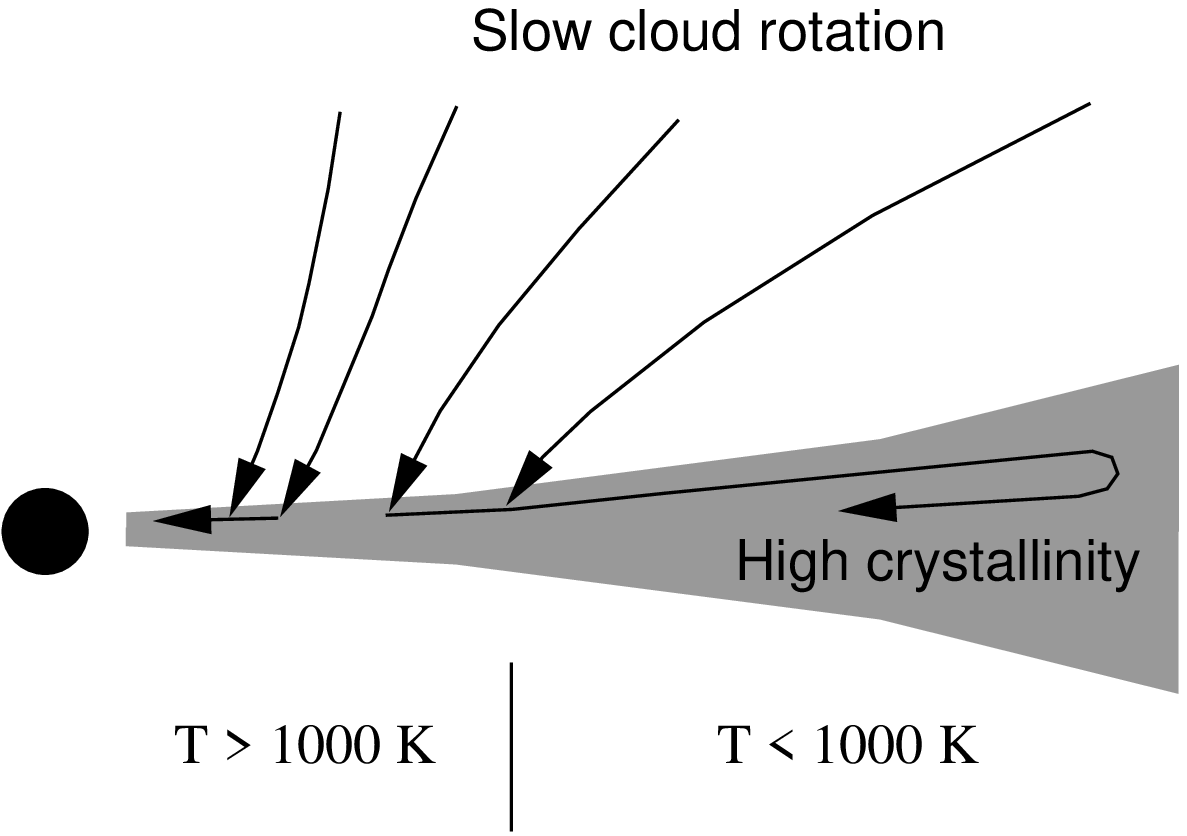}
\hspace{7em}
\includegraphics[width=5cm]{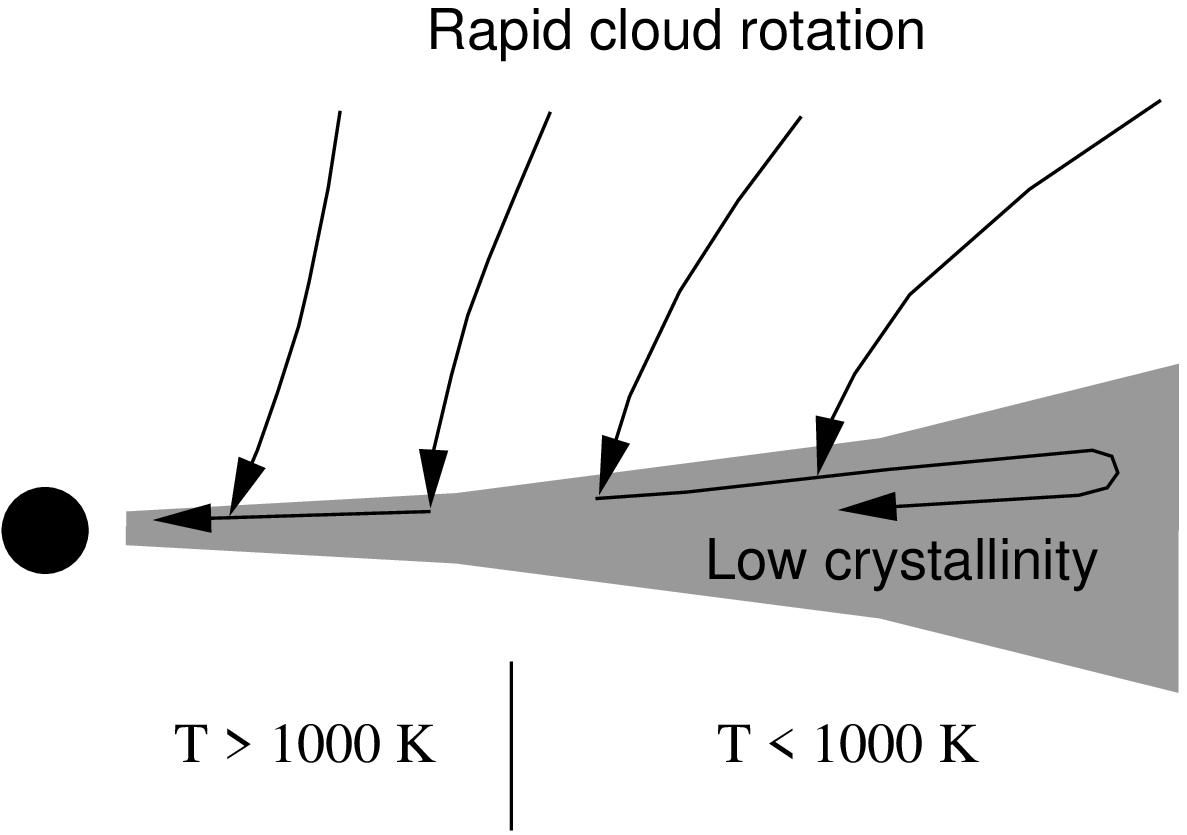}
}
\caption{\label{fig-cartoon} Cartoon of how the angular momentum of the
infalling matter affects the crystallinity of the resulting
disk. \revised{Left: low angular momentum infalling matter falls onto the
disk within a small region around the star. In this region the disk (in
particular in the early active phase) is hot enough to crystallize the
dust. Subsequent disk-expansion transports this material outward (some of
which later accretes back inward, Lynden-Bell \& Pringle
\citeyear{lyndenpring:1974}). Much (if not most) of the matter observed at
large radii originally came from these hot inner regions, and hence has a
high abundance of crystalline silicates. Right: high angular momentum matter
falls onto the disk at large radii, where the disk is cool. This matter will
not get crystallized. Even though radial mixing will pollute it partially
with crystalline silicates from smaller radii, the overall crystallinity of
the disk will then be much lower than in the high-angular-momentum case.}}
\end{figure*}

In this Letter we wish to revisit the radial mixing theory from a new
perspective. Instead of assuming a steady disk, or starting from a given
disk structure, we take a step back and start from the collapse of the
rotating pre-stellar cloud core. We assume that the collapse of such a core
leads to the formation of a {\em single} star and a disk, even though this
may not be the dominant mode of star formation. The properties and viscous
evolution of this disk are strongly influenced by the rotation rate of the
original cloud core (Nakamoto \& Nakagawa \citeyear{nakamotonakagawa:1994};
Hueso \& Guillot \citeyear{huesoguillot:2005}). Rapidly rotating clouds
deposit their mass onto the disk at large stellocentric radii, leading to 
\revised{the slower evolution of a more massive disk}
than if the cloud were rotating
slowly. Slowly rotating clouds deposit most material close to the star where
it will be strongly heated and then push outwards by viscous spreading,
influencing the thermal history of material found in the outer disk
(Fig.~\ref{fig-cartoon}).  Nakamoto \& Nakagawa
(\citeyear{nakamotonakagawa:1994}) and Hueso \& Guillot
(\citeyear{huesoguillot:2005}) have modeled the formation and viscous
evolution of protoplanetary disks using vertically integrated time-dependent
viscous disk models. Here we follow a similar approach but along with the
disk evolution we also passively evolve a dust population, including
a simple treatment of crystallization and radial mixing.

In this Letter we only demonstrate the principle and present first results.
In a follow-up paper we will describe the model in detail, explore the
parameter space, discuss all the physical uncertainties and caveats, and
compare the results to observational trends.

\section{Brief description of the model}
We start our simulations with a core with temperature $T_{\mathrm{bg}}$
(Kelvin), mass $M$ (gram), and solid-body rotation rate $\Omega$
(radian/second). For the sake of simplicity we assume the core to be a
singular isothermal sphere, required to apply the Shu collapse model.  Here
the collapse is triggered by an expansion wave propagating outward from the
center with the sound speed $c_s\equiv \sqrt{kT_{\mathrm{bg}}/\mu m_p}$,
initiating the collapse of every mass shell of the core it passes
through. The infall rate is constant: $\dot
M_{\mathrm{infall}}=0.975\;c_s^3/G$, which for our choice of temperature,
$T_{\mathrm{bg}}=15$~K, amounts to $\dot M_{\mathrm{infall}}=3\times
10^{-6}M_{\odot}/$year. For a cloud with a rotation rate $\Omega$ all the
mass will fall onto the equatorial plane {\em inward} of the centrifugal
radius $\rcentr{}\equiv \Omega^2R_{\mathrm{core}}^4 /GM_{\mathrm{core}}$,
where the radius of the core is
$R_{\mathrm{core}}=GM_{\mathrm{core}}/2c_s^2$. The way the infalling
material is distributed over the disk is described by Hueso \& Guillot
(\citeyear{huesoguillot:2005}).

The disk is described by the surface density $\Sigma$ as a function of
the radius $R$. It obeys the equation:
\begin{equation}
\frac{\partial \Sigma}{\partial t} +
\frac{1}{R}\frac{\partial R\Sigma v_R}{\partial R} = S
\comma
\end{equation}
where $S$ is the source function due to infalling matter onto the
disk. The azimuthal momentum equation yields an expression for the radial
velocity of the gas $v_R$:
\begin{equation}
v_R = - \frac{3}{\Sigma\sqrt{R}}\frac{\partial}{\partial R}\left(
\Sigma \nu \sqrt{R}\right)
\comma
\end{equation}
where $\nu$ is the viscosity coefficient, given by $\nu\equiv \alpha
kT_m/\mu m_p\Omega_K$. Here $\alpha$ is the parameter of viscosity (Shakura
\& Sunyaev \citeyear{shaksuny:1973}), which we take to be a global
parameter, $T_m$ is the midplane temperature, and $\Omega_K$ is the
Keplerian frequency. We also include an effective viscosity caused by
gravitational instabilities in regions where the Toomre parameter drops
below unity. The midplane temperature is determined by taking into account
heating due to viscosity and irradiation by the central star.

The dust is passively transported along with the gas, but is subject to
radial diffusion through turbulence and to crystallization in those regions
for which $T_{m}\gtrsim 800$~K. We therefore transport two dust components:
an amorphous one and a crystalline one, with surface densities $\Sigma_1$
and $\Sigma_2$, respectively. Crystallization transforms one into the other
at a rate given by the Arrhenius formula $\nu\exp(-T_c/T_m)\;\sec^{-1}$ with
the lattice vibration frequency $\nu=2\times 10^{13}$ (Fabian et
al.~\citeyear{fabianjaeger:2000}) and $T_c=38100\;$K (table 3 of
Bockele\'e-Morvan et al.~\citeyear{bockmorvan:2002}). The radial mixing
equation is:
\begin{equation}
\frac{\partial \Sigma_i}{\partial t} +
\frac{1}{R}\frac{\partial R\Sigma_i v_R}{\partial R} 
= \frac{1}{R}\frac{\partial}{\partial R}\left[R D \Sigma
\frac{\partial}{\partial R}\left(\frac{\Sigma_i}{\Sigma}\right)\right]
+ S_i
\end{equation}
where $D\equiv \nu/P_t$ is the diffusion coefficient, $P_t$ is a constant
that follows from numerical simulations (Johansen \& Klahr
\citeyear{johansenklahr:2005}; Carballido et
al.~\citeyear{carballido:2005}). The source term $S_i$ accounts for the
infall of new matter onto the disk (only feeding the amorphous component)
and for the transformation of amorphous silicates into crystalline ones. We
assume that the amorphous primordial dust survives the accretion shock on
the surface of the disk.

\section{Results}
As an example, we take a Herbig Ae/Be star with $M_{*}=2.5 M_{\odot}$,
$T_{*}=10,000$~K, $L=50~L_{\odot}$. The initial cloud is assumed to be at 15
K, and its rotation rate is taken to be $\Omega=1\times
10^{-14}\sec^{-1}$. We assume that the viscosity parameter is $\alpha=0.01$
throughout the disk.

\begin{figure}
\centerline{
\includegraphics[width=7cm]{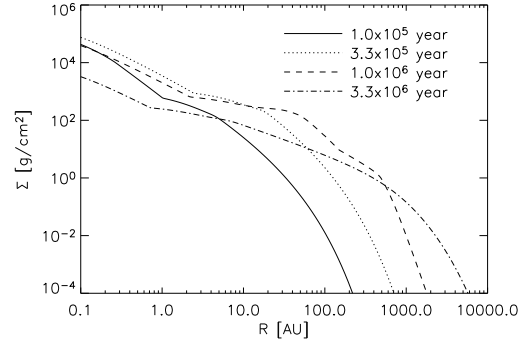}
}
\caption{\label{fig-example-hae-sigma} Surface density as a function of
radius in the disk, for different times after the onset of collapse of the
parent cloud.}
\end{figure}

In Fig.~\ref{fig-example-hae-sigma} the surface density profile as a
function of radius is shown at various epochs after the onset of collapse of
the parent cloud. In this example the centrifugal radius was located at 180
AU, i.e. all the matter from the infalling cloud fell within 180 AU from the
star onto the disk. Viscous friction then causes much of the mass to accrete
inward onto the star, while pushing a smaller fraction of the matter out to
very large radii in order to absorb the angular momentum of the inward
moving matter.

\begin{figure}
\centerline{
\includegraphics[width=7cm]{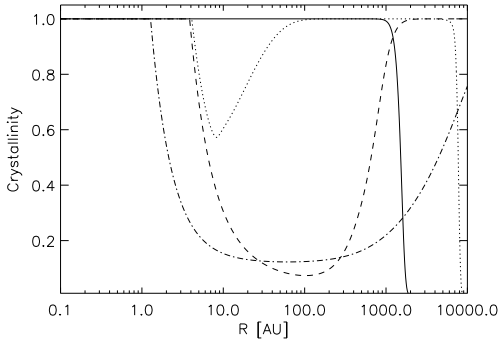}
}
\caption{\label{fig-example-hae-cryst} Abundance of 
crystalline silicates as a function of radius, for the same times as in 
Fig.~\ref{fig-example-hae-sigma}.}
\end{figure}

In the initial phases of the infall ($t\ll 10^6\;$year), however, the
centrifugal radius \revised{of the infalling matter} was smaller than 180 AU because the inner parts of the
rigidly rotating cloud had a lower specific angular momentum. Therefore, in
the early phases of the collapse most of the matter falls much closer to the
star than in the later stages. In these inner regions the disk becomes so
warm due to accretion that all the dust crystallizes or evaporates. As some
of this matter is pushed out for reasons of angular momentum conservations
(not necessarily radial mixing), this matter cools off and, if the dust was
evaporated initially, it recondenses \revised{directly into}
crystalline form. Hence, the spreading disk is transferring crystalline
matter from the inner disk to large radii, causing initially the disk to be
fully crystalline, as shown in
Fig.~\ref{fig-example-hae-cryst}.

In the later stages of the infall ($\sim 5\times 10^5\cdots 10^6\;$years),
the collapse wave reaches the outer shells of the cloud, triggering this to
fall onto the disk. Since this matter has a higher specific angular momentum
it falls onto the disk at larger radii, where the disk is cooler. This
matter therefore will stay amorphous and mix with the crystalline material
already in the disk, diluting the crystalline content.  The disk becomes
more amorphous again. The onset of this dilution with amorphous material is
seen in the $3.3\times 10^5$ year curve (dotted curve) in
Fig.~\ref{fig-example-hae-cryst}.  Dilution continues until infall ceases,
which is at slightly less than 1 Myr in this model.  Further viscous
spreading and radial mixing of the dust will cause the region beyond 10 AU
to have an almost constant crystallinity, while in between 1 and 10 AU
crystallinity declines with radius, in accordance with the radial mixing
theory. The level of the crystallinity in the outer (10-1000 AU) disk region
is a heritage from the infall phase set by the amout of cool
non-crystallized matter that was able to dilute the crystalline matter from
the very early infall phase.

\begin{figure}
\centerline{\includegraphics[width=9cm]{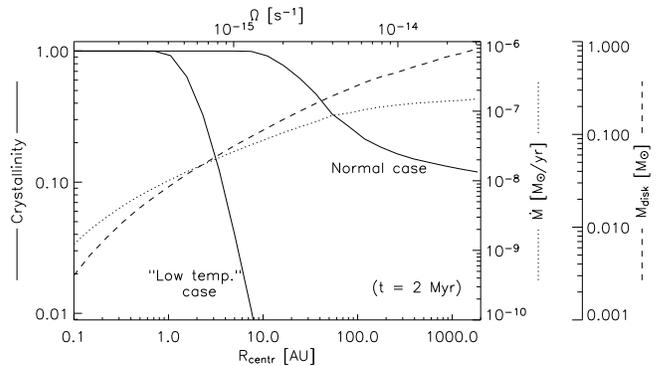}}
\caption{\label{fig-cryst-afo-omega} A number of disk properties plotted as
a function of cloud rotation rate $\Omega$, or equivalently as a function of
the centrifugal radius $\rcentr{}$. The latter is the fundamental variable
relevant for the disk evolution, but with the simple cloud collapse model it
translates in $\Omega$. The following quantities are plotted: the
crystallinity at 10 AU (i.e.~abundance of crystalline silicates:
\revised{black solid curves}), the mass accretion rate at the inner edge of
the disk (dotted curve), and the disk mass (dashed curve), all measured at 2
Myr after onset of collapse, for the model parameters described in the
text. \revised{The top black curve is for the model with the ``standard''
input physics. The bottom black curve is for the model in which the most
optimistically low midplane temperature estimate is used, and in which the
radial mixing efficiency is reduced by a factor of 10 (see text).}}
\end{figure}

In Fig.~\ref{fig-cryst-afo-omega} we plot the crystallinity at 10 AU at 2
Myr after the onset of collapse for a series of models with different cloud
rotation rate $\Omega$.  Also shown are the disk masses and accretion
rates at 2 Myr.

The figure shows that cloud rotation rates of at least $\Omega\gtrsim
4\times 10^{-15}\sec^{-1}$ are required to obtain crystallinity below 30\%,
which is the approximate upper limit obtained for a sample of Herbig Ae/Be
stars by van Boekel et al.~(\citeyear{vboekelmin:2005}).  Disks formed from
such rapidly rotating clouds are large and require long time to be accreted.
At 2 Myr they will still have high disk masses ($M_{\mathrm{disk}}\gtrsim
0.4 M_{\odot}$) and reasonably high accretion rate ($\dot M\gtrsim
10^{-7}\;M_{\odot}/$year).

The crystallinity depends strongly on the detailed treatment of the
temperature at the disk midplane. Our standard computation leads to rather
hot disks, but many effects might lower the disk temperature, including
convection and dust grain growth. The minimum temperature is the effective
temperature required to radiate away the energy input from both the
irradiation as well as the viscous heating. This lower temperature will
evidently yield lower crystallinity. Crystallinity can be suppressed even
further by reducing the radial mixing coefficient (i.e.~increasing the
Prandtl number $P_t$). The grey curve in Fig.~\ref{fig-cryst-afo-omega}
represents models with $P_t=10$ and minimum temperature. For these models a
level of 30\% crystallinity can be reached ($M_{\mathrm{disk}}=$0.05
$M_{\odot}$, $\dot M=10^{-8}M_{\odot}/$year at 2 Myr).

However, the above assumptions differ drastically from conventional
wisdom: The lowest possible estimate of the disk midplane temperature 
would require a Rosseland optical depth of roughly unity throughout the
disk, a rather unlikely coincidence.  The real temperature presumably lies
somewhere in between the two extremes we tested. The factor of 10 reduction
in radial mixing means that the angular momentum in the disk is much more
efficiently transported than passive trace elements in the disk. In
accretion disk theory the Prandtl number is typically assumed to be about
unity.

\section{Discussion and conclusion}
In this Letter we have shown that understanding the crystallinity of dust in
disks requires the modeling of the entire formation process of the star+disk
system, because much of the crystallization occurs in these very early
phases. \revised{Unlike the existing models in the literature where the
initially amorphous disks are gradually enriched with crystals via radial
mixing, our model predicts a major heating event and dust crystallization
{\em simultaneously} with the build-up of the accretion disk.  Thus, in
contrast to the gradual increase of crystallinity proposed by others, our
model predicts an initially high crystallinity, that possibly may even
decrease with time --- a prediction that might provide a natural explanation
for the probable age--crystallinity anti-correlation indicated by van Boekel
et al.~(\citeyear{vboekelmin:2005}) and Apai et
al.~\citeyear{apaiscience:2005}.}

The crystallinity, as well as many other disk parameters \revised{measured
after the main envelope-infall phase}, depends crucially on the angular
momentum of the infalling matter: low angular momentum leads to high
crystallinity and high angular momentum to low crystallinity. Our model only
includes thermal annealing as a crystallization process. Other processes
such as shock annealing (both within the disk itself and in the infall-shock
on the disk surface) are neglected. Nevertheless, our model typically
predicts high levels of crystallinity. A detailed comparison to observations
(e.g.~to Spitzer IRS spectra and/or VLT-MIDI interferometric spectroscopy
data) can therefore put strong constraints on the disk physics. The low
levels of crystallinity observed in some sources may require lowering the
vertical optical depth of the disk by dust coagulation.  We may also be
forced to reduce the radial mixing or modify other pieces of basic disk
physics. The study of the disk crystallinity, in combination with disk mass
and accretion rate, will therefore teach us about several aspects of
fundamental disk physics. In upcoming work we will study these aspects in
more detail and perform direct comparisons to observations.

\begin{acknowledgements}
We thank Carsten Dominik for his insightful input and careful reading of the
drafts. We also wish to thank Th.~Henning, H.-P.~Gail, E.~van Dishoeck,
P.~Myers, A.~Natta, M.~Walmsley and J.~Muzerolle for useful discussions.
This material is partly based upon work supported by NASA through the NASA
Astrobiology Institute under Cooperative Agreement No.~CAN-02-OSS-02.
Partial support for this work was provided by NASA through the contract
1268028 issued by the Jet Propulsion Laboratory, California Institute of
Technology under a contract with NASA.
\end{acknowledgements}

\bibliographystyle{apj}

\end{document}